\documentclass[%
pra,
twocolumn,
superscriptaddress,
 amsmath,amssymb,
 aps,
]{revtex4}

\usepackage{graphicx,subfigure}
\usepackage{float}
\usepackage{hyperref}
\usepackage{dcolumn}
\usepackage{bm,braket}
\usepackage{enumerate}
\usepackage{amsmath,amsfonts,amssymb,amsthm,bm,upgreek}
\usepackage[mathscr]{eucal}
\usepackage{dsfont}

\def\be{\begin{equation}}
\def\ee{\end{equation}}
\def\ba{\begin{eqnarray}}
\def\ea{\end{eqnarray}}

\begin{document}
\title{Quantum phase space with a basis of Wannier functions}
\author{Yuan Fang}
\affiliation{International Center for Quantum Materials, School of Physics, Peking University, Beijing 100871, China}
\author{Fan Wu}
\affiliation{International Center for Quantum Materials, School of Physics, Peking University, Beijing 100871, China}
\author{Biao Wu} \email{wubiao@pku.edu.cn}
      \affiliation{International Center for Quantum Materials, School of Physics, Peking University, Beijing 100871, China}
      \affiliation{Collaborative Innovation Center of Quantum Matter, Beijing 100871, China}
      \affiliation{Wilczek Quantum Center, School of Physics and Astronomy, Shanghai Jiao Tong University, Shanghai 200240, China}

\date{\today}

\begin{abstract}
A quantum phase space with Wannier basis is constructed: (i)
classical phase space is divided into Planck cells;  (ii) a complete set of Wannier functions are
constructed with the combination of Kohn's method and L\"owdin method such that each Wannier function
is localized at a Planck cell. With these Wannier functions one can  map a wave function unitarily 
onto phase space. Various examples are used to illustrate our method and compare it
to Wigner function. The advantage of our method is that it can smooth out the oscillations in wave functions 
without losing any information  and is potentially a better tool in studying quantum-classical correspondence. 
In addition, we point out that our method can be used for time-frequency analysis of signals. 
\end{abstract}

\maketitle

\section{\label{sec:level1}Introduction}
Phase space, where every point represents a  state in a classical dynamical system, 
is not only  a fundamental concept but also an important tool in classical mechanics. In contrast, 
in quantum mechanics, the fact that position and momentum operators are not commutative results 
in the difficulty of proper definition of quantum phase space.  Nevertheless, physicists have tried
various ways to adapt phase space into quantum mechanics. One famous  example 
is the reformulation of quantum mechanics in phase space with path integral by 
Feynman~\cite{Path}. 

Another well-known example is Wigner function, which gives a  representation of wave 
function in phase space~\cite{Wigner}. Later Husimi developed Q representation while 
Sudarshan and Glauber developed P representation  in phase space for wave function~\cite{Husimi,Sudarshan,Glauber}. 
However, all these three methods can only give us {\it quasi}-probability distributions in phase space:
The Wigner function and Sudarshan-Glauber P function can be negative; for the Husimi Q function, 
its marginal distribution for a pure state $\psi$ does not equal to $|\psi|^2$.  Despite these drawbacks, we
have seen tremendous developments of all these methods over the years because they are natural 
bridges between quantum and classical dynamics and also have many practical applications in 
quantum optics, nuclear and particle physics, condensed matter and mesoscopic systems~\cite{PhaseBook,WignerMeasure}.
 
In 1929 von Neumann already suggested a different way to map a wave function onto 
phase space~\cite{Von1929,vonNeumann2010qhtheorem}.  In von Neumann's method, 
one divides classical phase space into Planck cells and then finds a set of orthonormal wave functions 
which are localized at these Planck cells.  With these orthonormal wave functions 
served as a basis, a wave function is mapped unitarily to phase space, and the amplitudes 
at  the Planck cells give  us a true probability distribution. 
Von Neumann's  motivation was to establish quantum phase space so that he could
borrow many ideas from classical statistical physics to set up a foundational framework
for quantum statistical physics~\cite{Von1929,vonNeumann2010qhtheorem}. 

However, von Neumann only showed how these localized 
wave functions could be found in principle but did not offer an efficient approach to compute them.  
Von Neumann's method was developed in a recent work~\cite{Han} where Kohn's 
orthogonalization method~\cite{Kohn73} was employed and a set of Wannier functions 
localized at Planck cells were found.

 In this work we further develop this quantum phase space with Wannier functions as its basis.
 We find a more efficient approach to compute these Wannier functions  by using 
 L\"owdin's orthogonalization method~\cite{Lowdin,Aiken} on top of the 
 Kohn's method~\cite{Kohn73}. With this Wannier basis, a wave function can be mapped unitarily
 onto phase space.  The amplitude at each Planck cell is complex in general and, however,
 due to the unitarity of this mapping,  the square of its amplitude magnitude is {\it true} probability. 
 This is the crucial difference between our method and well-known Wigner, P or Q function. The latter
can only give us quasi-probability. With our method, it is now possible to test numerically many 
fundamental ideas proposed by  von Neumann in 1929 ~\cite{Von1929,vonNeumann2010qhtheorem}.

 Using various concrete examples, we compare our unitary mapping to Wigner function. There are  
 two key features in the comparison. ({\it i}) Our unitary mapping is very effective to smooth out the oscillations 
 in a wave function and produces a probability distribution that in some cases resembles a classical trajectory 
 while the Wigner function can not.  ({\it ii}) The Wigner function is coarse-grained by averaging over Planck cells;
 the resulting distribution is very different from the original Wigner function but 
 very similar to the true probability distribution obtained with our method. This shows that one can roughly 
 get a probability distribution in phase space by coarse-graining 
 Wigner function. However,  a lot of  information is lost to with  coarse-graining whereas our unitary 
 mapping  does not lose any information.  Such a comparison shows that our unitary mapping 
 can be an excellent tool for studying the quantum-classical correspondence, the central theme 
 of quantum chaos~\cite{StockmannBook}. In the end, we further point out that our method can be used
 for time-frequency signal analysis. 
 
The rest of this paper is organized as follows. In Section II,  von Neumann's method is reviewed 
along with the work in Ref.~\cite{Han}. In Section III, our method is described in detail. We then
discuss how localized our Wannier functions are in Section IV. In Section V,  we compare our
unitary projection to Wigner function with various examples.  In Section VI, we use
an example to show how our method can be applied to signal analysis. In the end we draw some conclusions
in Section VII.

\section{Review of Von Neumann's Method}
Von Neumann in 1929 suggested a method to construct quantum phase space~\cite{Von1929,vonNeumann2010qhtheorem}, 
which consists with two steps: 
({\it i}) dividing the classical
phase space into Planck cells; ({\it ii}) finding a set of orthonormal  wave functions which are localized at Planck cells. 
Von Neumann suggested to find these orthonormal wave functions by  orthogonalizing a set of Gaussian 
wave packets of width $\zeta$ with the Schmidt method,
\begin{equation}
\label{g}
g_{j_x,j_k} \equiv \exp\left[-\frac{(x-j_xx_0)^2}{4\zeta^2}+ij_kk_0x\right]
\end{equation}
where $j_x$ and $j_k$ are integers. When $x_0k_0=2\pi$, this set of Gaussian packets is  complete.

This method shows that in principle one can map a wave function unitarily onto phase space and establish
a probability distribution for a quantum state in phase space. This proof of principle was enough 
for von Neumann to prove in an abstract way of quantum ergodic theorem and 
quantum H theorem~\cite{Von1929,vonNeumann2010qhtheorem}.

However, von Neumann's construction is not very computationally practical due to the following drawbacks:
 \begin{enumerate}[i)]
 \item The Schmidt orthogonalization procedure is computationally costly, rendering it 
 numerically not feasible. 
 \item The  wave functions constructed in this way lack spatial translational symmetry. 
 Since there should be no difference of measuring coordinates at different sites, such a symmetry is desired.
 \item This method is very sensitive to the order of the orthogonalization procedure and will produce base 
 functions with large tails (or standard deviations) bearing little resemblance to the original Gaussian functions.
\end{enumerate}

In Ref.~\cite{Han}, Han and Wu were able to remove the second drawback. In their approach, the subscript $j_k$ is treated
as band index and the Gaussian wave packets with the same $j_k$ are orthonormalized with  Khon's approach~\cite{Kohn73} 
to become a set of Wannier functions  whose spacial translational symmetry is guaranteed. At the same time, 
the computational cost is reduced substantially such that the whole method is now computationally feasible. 
However, the orthogonalization among Gaussian wave functions with different $j_k$ is still Schmidt and
the third drawback remains. In this work we employ L\"{o}wdin orthogonalization 
method~\cite{Lowdin} on top of Kohn's method.  As the L\"{o}wdin orthogonalization produces
a set of orthonormal vectors which are the most faithful to the original non-orthogonal vectors~\cite{Aiken}, 
the orthogonalization result is unique and independent of order of orthogonalization. So the third drawback is removed. 
Furthermore, the L\"{o}wdin method is more efficient and can reduce the computational cost dramatically. 

In Ref.~\cite{Von1929,vonNeumann2010qhtheorem}, von Neumann proposed many fundamental ideas using
his quantum phase space; however, these ideas had remained on the abstract level before our work. 
For example, von Neumann defined an entropy for pure quantum states using the probability distribution 
in his quantum phase space. However, there was no practical way to compute this entropy. 
With our method,  we can now compute such an entropy numerically~\cite{Han}.

\section{Our Method}
We focus on two dimensional phase spaces. Generalization to higher dimensions
are straightforward as done in Ref.\cite{Han}.  The detailed procedure of our method 
is elaborated as follows.
\begin{enumerate}[i)]
  \item  Choose an initial set of localized wave packets such as the Gaussian wave packets $g_{j_k}(x)\equiv g_{0,j_k}(x)$ in Eq.\eqref{g}. Find their Fourier transform $\tilde{g}_{j_k}(k)\equiv \mathcal{F}\{g_{j_k}(x)\}$. In our calculations, we choose $x_0=1,k_0=2\pi,\zeta=(2\pi)^{-1}$, and we have
      \begin{equation}
      \tilde{g}_{j_k}(k) \equiv \exp\Big[-(\frac{k}{2\pi}-j_k)^2\Big]\,.
      \end{equation}
  \item At a fixed $k\in [0,2\pi)$, for each $j_k$, we construct a columnwise vector whose $n$th element is $\tilde{g}_{j_k}(k+2n\pi)$.  We denote this vector 
 by $f_{k,j_k}$.
      \begin{equation}
      f_{k,j_k}=[\tilde{g}_{j_k}(k+2n\pi)]^T_{n\in\mathds{Z}}\,,
      \end{equation}
      where the superscript $T$ represents transpose. In numerical calculation, one needs to choose a cut-off $N$
      so that $-N\le n\le N$. These vectors
      $f_{k,j_k}$ with different $j_k$ are not orthogonal to each other. 
  \item Apply L\"{o}wdin orthogonalization to $f_{k,j_k}$:  (a) put these vectors together to form a matrix
      \begin{equation}
      F\equiv[ f_{k,-J_k},\cdots,f_{k,j_k}, f_{k,(j_k+1)},\cdots,f_{k,J_k}]
      \end{equation}
      where $J_k$ is the cut-off for $j_k$; (b) the matrix of the orthonormalized vectors is
      \begin{eqnarray}
      &&[u_{k,-J_k},\cdots,u_{k,j_k}, u_{k,(j_k+1)},\cdots,u_{k,J_k}] \nonumber\\
      &=&F (F^\dagger F)^{-\frac12};
      \end{eqnarray}
     (c) let $\tilde{w}_{j_k}(k+2n\pi) = u_{k,j_k}(n)/\sqrt{2\pi}$.
  \item The interval $[0,2\pi)$ is evenly divided into $N_k$ points in numerical calculations. 
  For every k of these $N_k$ points, repeat step (\romannumeral2) and step (\romannumeral3). Finally, after Fourier transform, we get a set of orthonormal basis of Wannier functions $\{w_{j_x,j_k}\}$, 
      \begin{equation}
      w_{j_x,j_k}(x)\equiv w_{j_k}(x-j_x)\,.
      \end{equation}
      For simplicity, from now on we will adopt Dirac notation and  let $\ket{w_j}=\ket{w_{j_x,j_k}}$, where 
      $(j_x,j_k)$ is simplified to $j$ whenever no confusion arises.  Note that in our calculation we set $N_k=2J_k$. 
\end{enumerate}
For these wave functions $\ket{w_j}$, the relation $\braket{w_{j_x,j_k}|w_{j_x,j'_k}}=\delta_{j_k,j'_k}$ is
guaranteed explicitly by the L\"owdin orthogonalization in the above procedure and the relation $\braket{w_{j_x,j_k}|w_{j'_x,j_k}}=\delta_{j_x,j'_x}$ is guaranteed implicitly by Kohn's method.  The full orthonormal relation 
$\braket{w_{j}|w_{j'}}=\delta_{j,j'}$ is then achieved.  Note that for a given set of non-orthogonal 
vector  the L\"owdin orthogonalization produces a unique set of 
orthonormal vectors~\cite{Lowdin,Aiken}. In contrast, the result of the Schmidt orthogonalization depends on the order of 
orthogonalization. As a result, our procedure gives rise to a unique set of Wannier functions once
the initial trial wave function, such as the Gaussians in Eq.\eqref{g}, are given. This removes the
third drawback in von Neumann's method. At the same time, it reduces further the computational cost. 

As  pointed out  above, although they are not orthonormal, 
the Gaussian wave functions in Eq.\eqref{g} are a complete set of basis when $x_0k_0=2\pi$. This is
already implicitly mentioned by von Neumann~\cite{Von1929,vonNeumann2010qhtheorem}. 
Consequently, the Wannier functions constructed out
of these Gaussian functions with our method form a complete set of orthonormal basis. This means that
the volume of a Planck cell is $x_0p_0=x_0\hbar k_0=2\pi\hbar=h$ with $h$ being the Planck constant. 

We summarize the basic feature of our quantum phase space: (1) It is made of Planck cells; (2) each 
Planck cell is assigned a Wannier function $\ket{w_j}$; (3) all the Wannier functions form a complete set of orthonormal 
basis.  Any given wave function $\ket{\psi}$ can now be mapped  onto our quantum phase space
\be
\label{mapping}
\ket{\psi}=\sum_j \ket{w_j}\braket{w_j|\psi}\,.
\ee
We emphasize that this mapping is linear and unitary, which is different from Wigner function, P representation,
or Q representation that are nonlinear and not unitary.  As a result, $p_j=|\braket{w_j|\psi}|^2$ is the probability 
at Planck cell $j$,  a true probability distribution over phase space.

\begin{figure}[ht]
  \centering

  	 \includegraphics[width=8cm]{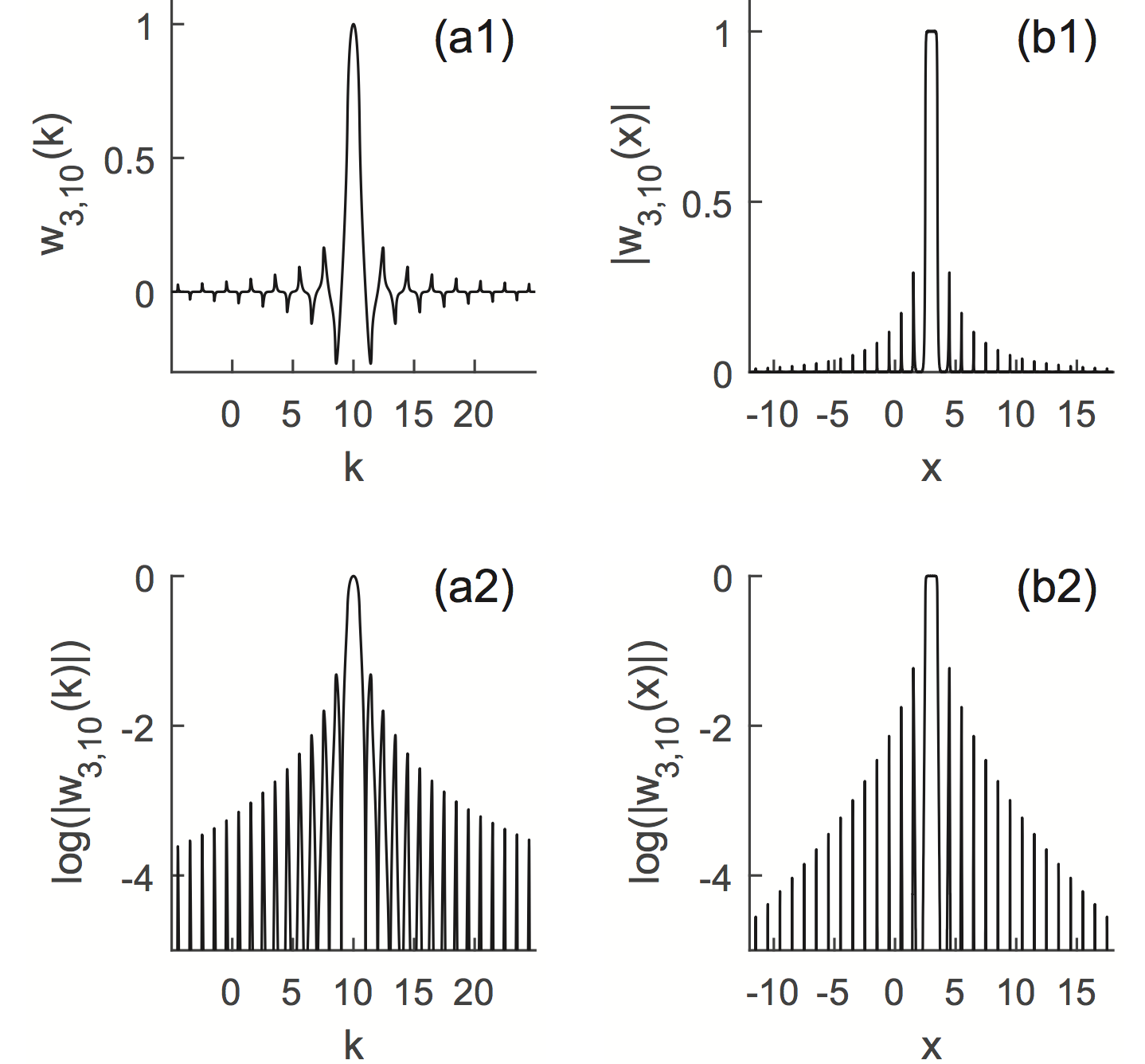}
  
 \caption{(a1,a2)
 Wannier function $w_{3,10}$ in the $k$ space; (b1,b2)  Wannier function $w_{3,10}$ in the $x$ space.  
 In $k$ space, the Wannier function is real and plotted directly; in $x$ space, the Wannier function
 is complex and its amplitude is plotted. 
 In the lower two panels,  the amplitude of $w_{3,10}$ is plotted in the semi-log format, 
 showing exponentially decaying tails in both $x$ and $k$ spaces.  The unit of $x$ is $x_0$ and
 the unit of $k$ is $k_0$. In our calculation, $J_k=40$, $N_k=80$,  and $N=50$.  }
 \label{fig1}
\end{figure}

\section{Localization of Wannier Functions}
In this section we examine how localized the above Wannier functions are. Shown in Fig.\ref{fig1} is one typical Wannier function $w_{3,10}$ in both $k$ space and $x$ space. This Wannier function is localized near 
the site $(x=3,\quad k=10\times2\pi)$ and is obtained by choosing $J_k=40$, $N_k=80$,  and $N=50$ with our method. 
It is clear from the two lower panels which are the semi-log plots of their counterparts in the upper panels that
the Wannier function is exponentially localized in both $k$ space and $x$ space.


\begin{figure}[ht]
  \centering
 \includegraphics[width=8cm]{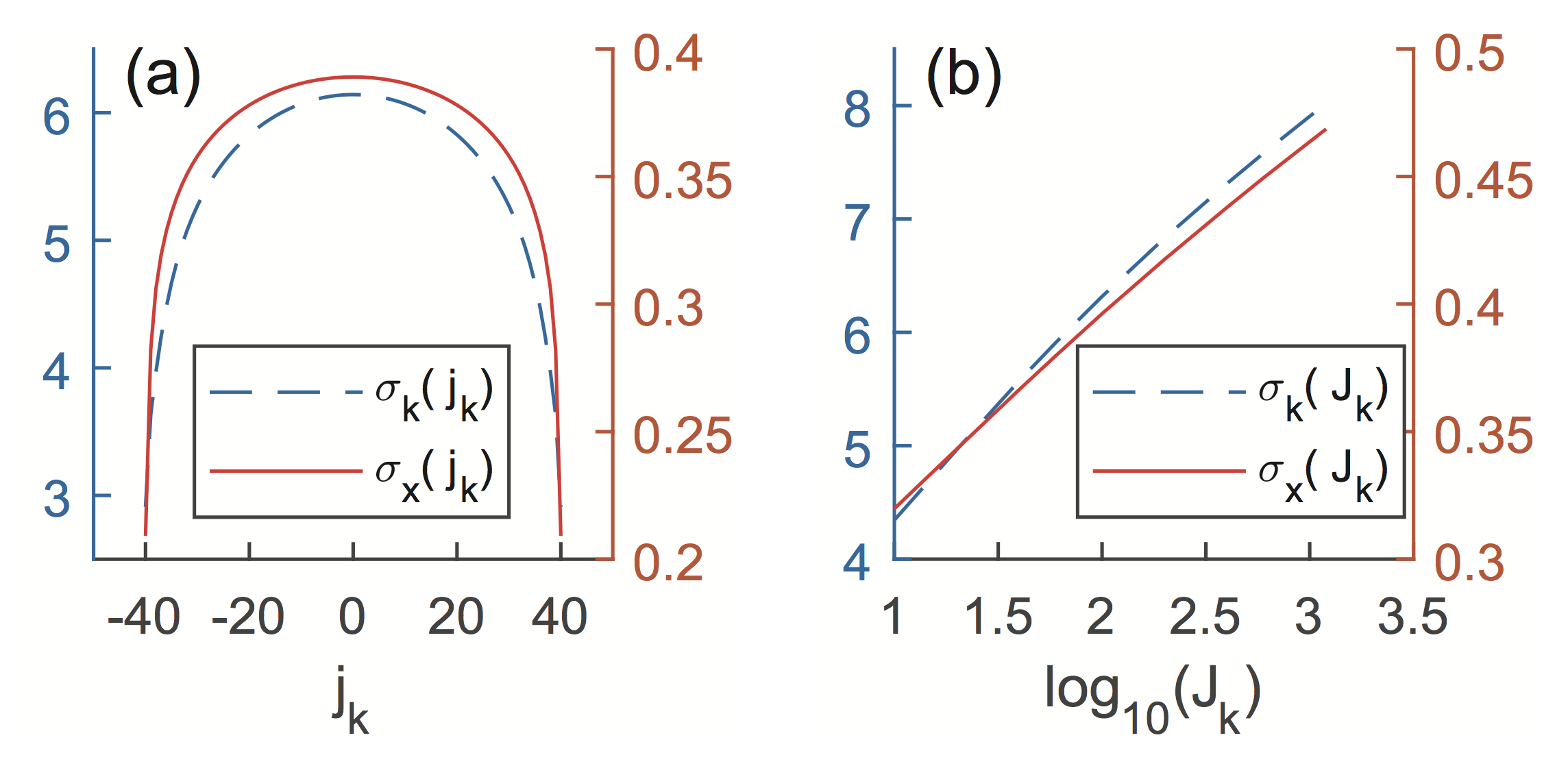}
 \caption{ (color online) (a) The widths $\sigma_x$ (red) and $\sigma_k$ (blue) of Wannier functions in both $x$ and $k$ spaces 
 as functions of $j_k$. $J_k=40$, $N_k=80$,  and $N=50$. 
 (b) The widths $\sigma_x$  and $\sigma_k$  at $j_k=0$ as function of $J_k$, 
 the cut-off of $j_k$. $N_k=2J_k$ and $N=50$.}
 \label{sigma}
\end{figure}

To measure the localization quantitatively, we use the standard deviation that is defined as
\begin{equation}
\sigma_x(j) = \langle w_j|(x-\langle x\rangle_j)^2|w_j\rangle^\frac12
\end{equation}
\begin{equation}
\sigma_k(j) = \langle w_j|(k-\langle k\rangle_j)^2|w_j\rangle^\frac12
\end{equation}
As Wannier functions have translational symmetry with respect to $j_x$, which is ensured by Khon's method, 
we only need to consider Wannier functions $w_{0,j_k}$.  Fig.\ref{sigma}(a)  illustrates how the standard deviations 
of these Wannier functions change with $j_k$. The figure shows that both $\sigma_x$ and $\sigma_k$ have maximal values
at $j_k=0$ and then decrease monotonically when $|j_k|$ increases. At the two ends where $|j_k|$ is large, 
we roughly have $\sigma_x\cdot\sigma_k \sim 0.5$, the lower limit of any wave packet demanded by the uncertainty relation. 
Both $\sigma_x$ and $\sigma_k$ are much larger, consequently, $\sigma_x\cdot\sigma_k$ is much larger than 0.5
when $|j_k|$ is small. 

When the size of quantum phase space in our numerical calculation, i.e., the cut-off $J_k$ increases, 
we find that the maximal values of both $\sigma_x$ and $\sigma_k$ increase. These two monotonic relations are 
plotted in  Fig.\ref{sigma}(b). The data in the figure apparently show that the increase of both $\sigma_x(j_k=0)$ and 
$\sigma_k(j_k=0)$  with $J_k$ is sub-logrithmic. This suggests that both 
$\sigma_x$ and  $\sigma_k$ may have finite upper limits.  Bourgain proved that such a basis exists 
in principle~\cite{bourgain1988remark}. However, Bourgain's approach of construction also involves 
Schmidt orthogonalization, which is computationally expensive.

\section{Comparison with Wigner function}
Our construction of quantum phase space yields a natural way to map a wave function onto phase space
as in Eq.\eqref{mapping}. Here we compare it to existing  methods that
map a wave function onto phase space. We focus on Wigner function as it is the most 
widely used method~\cite{PhaseBook}.  The differences between Wigner function
and our method are obvious: (i) Wigner function is a nonlinear mapping while ours is a linear
unitary mapping; (ii) Wigner function is real and continuous in phase space while ours produces
a discrete and complex function in phase space. However, as we will see, they bear some similarity
after Wigner function is coarse-grained.

Wigner function (or Weyl transform of density matrix) of a wave function $\psi(x)$ is defined as~\cite{Wigner,PhaseBook}
\begin{equation}
W(x,k) = \frac{1}{\pi}\int\psi^*(x+y)\psi(x-y)e^{2iky}dy\,.
\end{equation}
The Wigner function can be coarse-grained with a function $h(x,k)$ that is localized in phase space
\begin{equation}
W_h(x,k) = \int h(x'-x,k'-k)W(x',k') dx' dk'\,.
\end{equation}
The function $h(x,k)$ is usually chosen to be localized at a Planck cell. One popular choice is 
$h(x,k) = \frac 1\pi \exp({-\frac {x^2}{\sigma_x^2} -\frac {k^2}{\sigma_k^2}})$ 
with $\sigma_x\sigma_k=\frac 12$~\cite{PhaseBook}. In our calculation, we choose 
\begin{eqnarray}
h(x,k) &=& \left[H(x+\frac12)-H(x-\frac12)\right] \nonumber\\
&&\times\left[H(k+\frac12)-H(k-\frac12)\right]\,,
\label{hc}
\end{eqnarray}
where $H(x)$ is the Heaviside function. This $h$ function is intuitively simple as it facilitates an integration
precisely over a Planck cell. We shall use some typical and simple 
examples to compare our unitary projection to Wigner function. 

\begin{figure}[t]
  \centering
 \includegraphics[width=8cm]{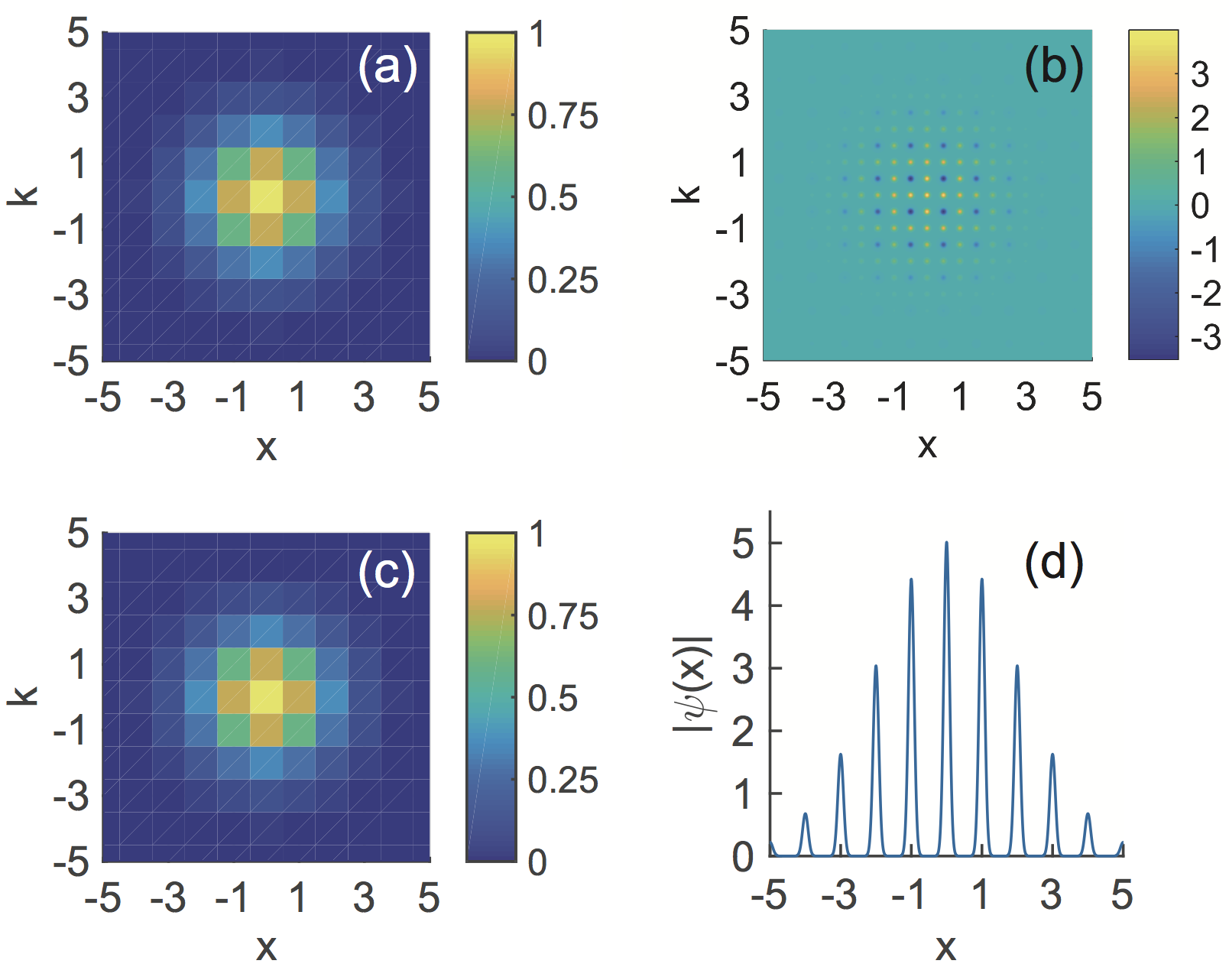} \\
 \caption{(color online) (a) Gaussian packet in our quantum phase space with the Wannier basis;  
   (b) the corresponding Wigner function, where the negative parts are surrounded by positive parts; 
  (c) the discrete coarse-grained Wigner function; (d) the amplitude in the $x$ space. The unit of $x$ is $x_0$ and
 the unit of $k$ is $k_0$.
}
 \label{gaussian}
\end{figure}

\subsection{Gaussian packet in our quantum phase space}
As the first example, we consider the following wave function
\begin{equation}
\label{gauf}
\psi(x) =  \sum \limits_{j_x,j_k} w_{j_x,j_k}(x)e^{-j_x^2-j_k^2}\,.
\end{equation}
This wave function can be regarded as a discrete Gaussian packet in our quantum phase 
space as shown in Fig.\ref{gaussian}(a).  It is positive in every Planck cell.  In contrast, 
as seen in Fig.\ref{gaussian}(b),  its corresponding Wigner function has either positive or negative values. 
Interestingly, this Wigner function becomes significantly different from zero only at integer or half integer coordinates. 
Its negative spots at half integer coordinates are   surrounded by positive spots at integer coordinates. 
This is a  reflection of the oscillations of the wave function $\psi(x)$ in the $x$ space (see Fig.\ref{gaussian}(d)). 

The coarse-grained Wigner function is plotted in Fig.\ref{gaussian}(c). Its overall feature looks very similar to 
Fig.\ref{gaussian}(a). Despite this similarity, we need to bear in mind that this coarse-grained Wigner function
does not give probability at a given Planck cell. Later we will show an example, where the coarse-grained Wigner function
can still be negative at some Planck cells. By coarse-graining, some information is lost while the projection 
with our Wannier basis is unitary and no information is lost. 

As any smooth function can be roughly regarded as a superposition of many Gaussian functions,  what we observe
from this typical example is quite general: ({\it i}) Our unitary 
projection onto the quantum phase space is very effective in smoothing out the oscillations in wave function while
Wigner function is not. ({\it ii}) Wigner function gives us only quasi-probability; it is still a quasi-probability when
averaged over a finite region such as a Planck cell. However, from the similarity between the  coarse-grained 
Wigner function and our unitary projection (Fig. \ref{gaussian}(b) vs. (c)), one can conclude that
the quasi-probability distribution of a coarse-grained Wigner function can be roughly regarded 
as a true probability distribution. In a sense, one can not claim this with great confidence before our work:
there was no unitary mapping to phase space before our work and therefore no true probability distribution; 
without comparison to a true probability distribution, one would not know how close a coarse-grained Wigner function
is to a true probability distribution. As our method can give arise to a true probability distribution, 
one is then  allowed to use it to define an entropy for pure quantum states~\cite{Von1929,vonNeumann2010qhtheorem,Han} 
One can not use Wigner function with or without coarse-graining for this purpose. 


\begin{figure}[t!]
  \centering
 \includegraphics[width=8cm]{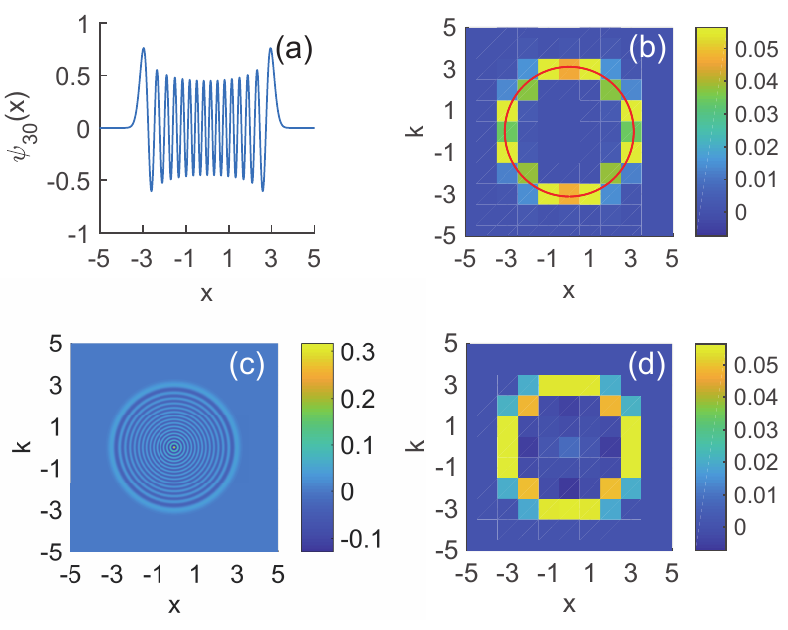}
 \caption{ (color online) The 30th energy eigenstate of an harmonic oscillator. 
(a) Its wave function in the $x$ space; (b) its unitary projection onto our quantum phase space (the red circle
is the corresponding classical trajectory); 
 (c) its Wigner function; 
 (d) its discrete coarse-grained Wigner function. The unit of $x$ is $x_0$ and
 the unit of $k$ is $k_0$.
 }
 \label{ho}
\end{figure}

\subsection{Harmonic Oscillator}

Harmonic Oscillator is one of the simplest problems in quantum mechanics. Its $n$th energy eigenfunction is
\begin{equation}
\psi_n(x) = (\frac{1}{2^nn!})^{1/2}\pi^{-1/4}\exp(-x^2/2)H_n(x)\,,
\end{equation}
where $H_n(x)$ are the Hermite polynomials. We choose  $n = 30$. The wave function of $\psi_{30}$ in the $x$ space 
is shown in Fig.\ref{ho}(a). And its unitary projection onto our quantum phase space is shown in Fig.\ref{ho}(b) where
we see that the wave function has significant weights along the classical orbit. As in the first example, 
the oscillations in $\psi_{30}(x)$ have disappeared in the Wannier representation. Looking more carefully,  one can find
that the weights  near grids $(3,0)$ and $(0,3)$ are a slightly different. It is because our Wannier basis does not have translational symmetry in the $k$ direction.

One very important feature in Fig.\ref{ho}(b) is that most of the probabilities concentrate along a circle, which is the
corresponding classical trajectory. This is not the case for the corresponding Wigner function. 
The Wigner function is shown on Fig.\ref{ho}(c) where we can see  15 circles, a reflection of 
the oscillations in $\psi_{30}(x)$. The center has the largest density of distribution. Therefore, 
our unitary projection can produce a probability distribution that resembles a classical trajectory while
Wigner function can not. In fact, we have applied our method to more sophisticated systems, 
where the quantum probability distribution in phase space obtained with our method bears strikingly similarity
to its classical ensemble distribution in phase space. Unfortunately, these results are beyond this work and
will be presented elsewhere. This shows that our unitary projection is a better tool to establish 
quantum-classical correspondence, the central subject in quantum chaos~\cite{StockmannBook}.

Fig.\ref{ho}(d) illustrates 
the coarse-grained Wigner function whose overall features look quite similar to Fig.\ref{ho}(b). 
However, there are minor differences, for example, it is symmetric in both $x$ and $k$ directions and it is
positive at the center $(0,0)$. 
We note one important feature:  the coarse-grained Wigner function is negative at  $(\pm2,0)$ and $(0,\pm2)$. 
This shows that the coarse graining with the chosen $h$ function  in Eq. \eqref{hc} does not guarantee 
positive value at a given Planck cell. Despite its similarity to our unitary projection, 
a coarse-grained Wigner function is still a quasi-probability.

\begin{figure}[t]
  \centering
\includegraphics[width=8cm]{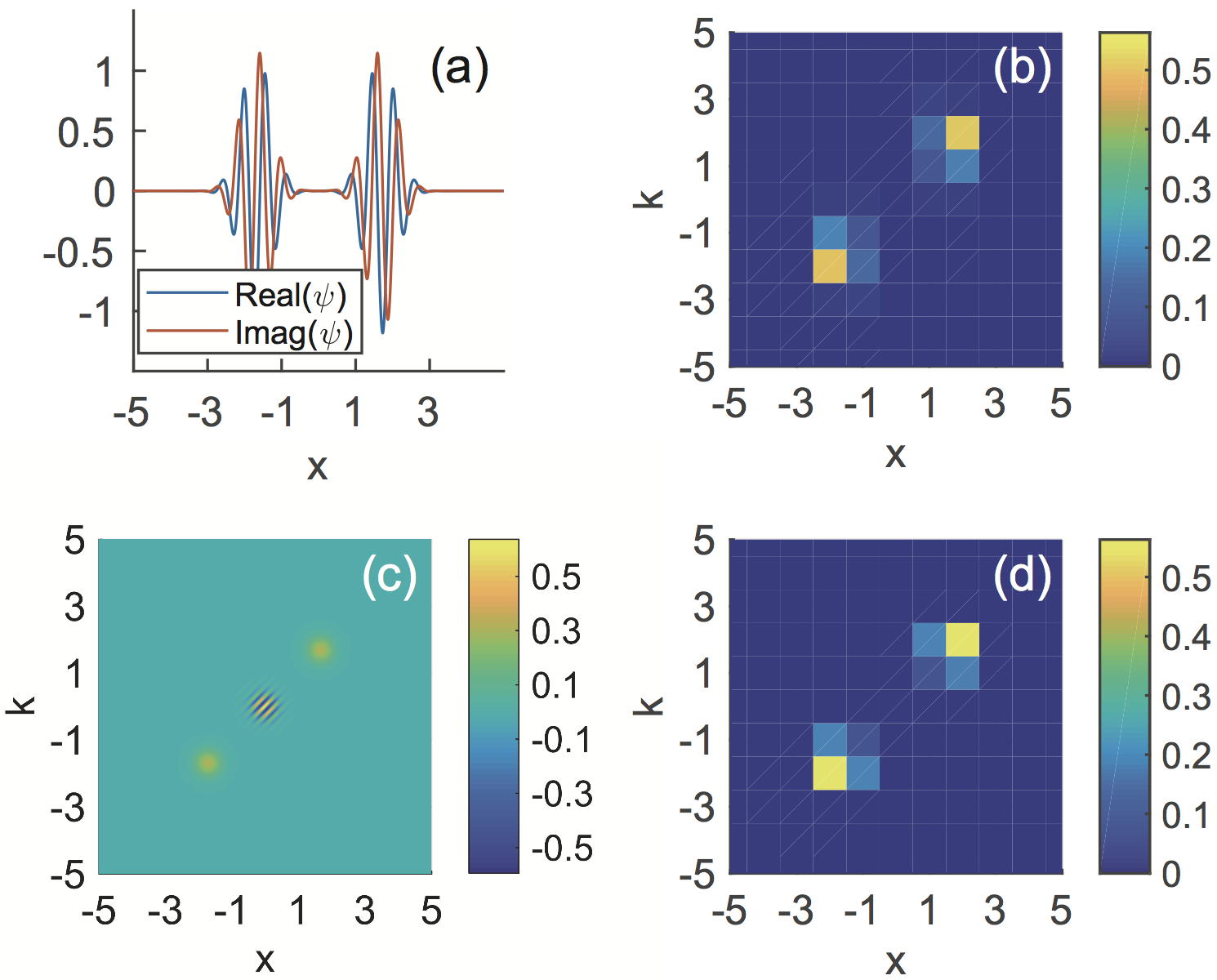} \\
 \caption{(color online) 
 Sch\"{o}rdinger cat state $|{\rm cat}\rangle= |\alpha\rangle+|-\alpha\rangle$ with 
 $\alpha = 3+3i\rangle$. 
 (a) Its wave function in the $x$ space; 
 (b) its unitary projection onto our quantum phase space; 
 (c)  its Wigner function;
 (d) its discrete coarse-grained Wigner function. The unit of $x$ is $x_0$ and
 the unit of $k$ is $k_0$.
}
 \label{cat}
\end{figure}

\subsection{Sch\"{o}rdinger cat state}

In quantum optics, a cat state is defined as the superposition of two coherent states with opposite phase:
\begin{equation}
|{\rm cat}\rangle = |\alpha\rangle+|-\alpha\rangle = 2e^{\frac{-|\alpha|^2}{2}}\sum\limits_n\frac{\alpha^{2n}}{\sqrt{(2n)!}}|2n\rangle\,,
\end{equation}
where $|2n\rangle$ is a Fock state with $2n$ particles. In our calculation,  we choose $\alpha = 3+3i$.  
The wave function in the $x$ space is shown on Fig.\ref{cat}(a). The wave function looks like two moving Gaussian packets localized near $x=-2$ and $x = 2$. Its unitary projection onto the quantum phase space is shown on Fig.\ref{cat}(b). 
The wave function looks again rather smooth in the quantum phase space and is localized around two regions.  

Its  Wigner function is plotted in  Fig.\ref{cat} (c) and it has a rapidly oscillating center, which is 
regarded as an iconic feature of a coherent cat state~\cite{TilmaPRA}. 
 However, this oscillating center disappears in the coarse-grained Wigner function  in Fig.\ref{cat}(d) and
in our unitary projection in Fig.\ref{cat}(b). This means that the probability around this center 
is in fact close to zero. 

With the examples above, we can conclude that our unitary projection of a wave function onto 
quantum phase space with the Wannier basis produces a result looking very similar to the coarse-grained
Wigner function. This has two implications: ({\it i}) 
Our unitary project is very effective to smooth out the oscillations in a wave function
However, our projection is unitary and does not lose any information while
a lot of information is lost in the coarse-graining.  ({\it ii}) As a result,  our unitary project
can produce a true probability distribution resembling a classical trajectory as most dramatically
seen with the example of harmonic oscillator.  The oscillations between positive and negative values
in Wigner functions (see Figs.\ref{gaussian}(b),  \ref{ho}(c)\&\ref{cat}(c)) are regarded as an indication 
of  ``quantumness" in the quantum state~\cite{PhaseBook,TilmaPRA}. However, this also makes it difficult to build 
a connection between quantum dynamics and classical dynamics in phase space. One way to go around
this difficulty is to remove some information of a wave function, i.e., coarse-graining.  
Our unitary projection can achieve this goal without losing any information. The reason is that 
the oscillations are hidden in the Wannier basis. Therefore, our unitary projection is a 
better tool in studying the quantum-classical correspondence in phase space. It will also 
be interesting to compare our method to Wigner function in other applications such as in quantum optics~\cite{WignerMeasure}. 
This is beyond the scope of this work. 

\section{Application in Time-frequency Signal Analysis}
\begin{figure}[!t]
  \centering
\includegraphics[width=8cm]{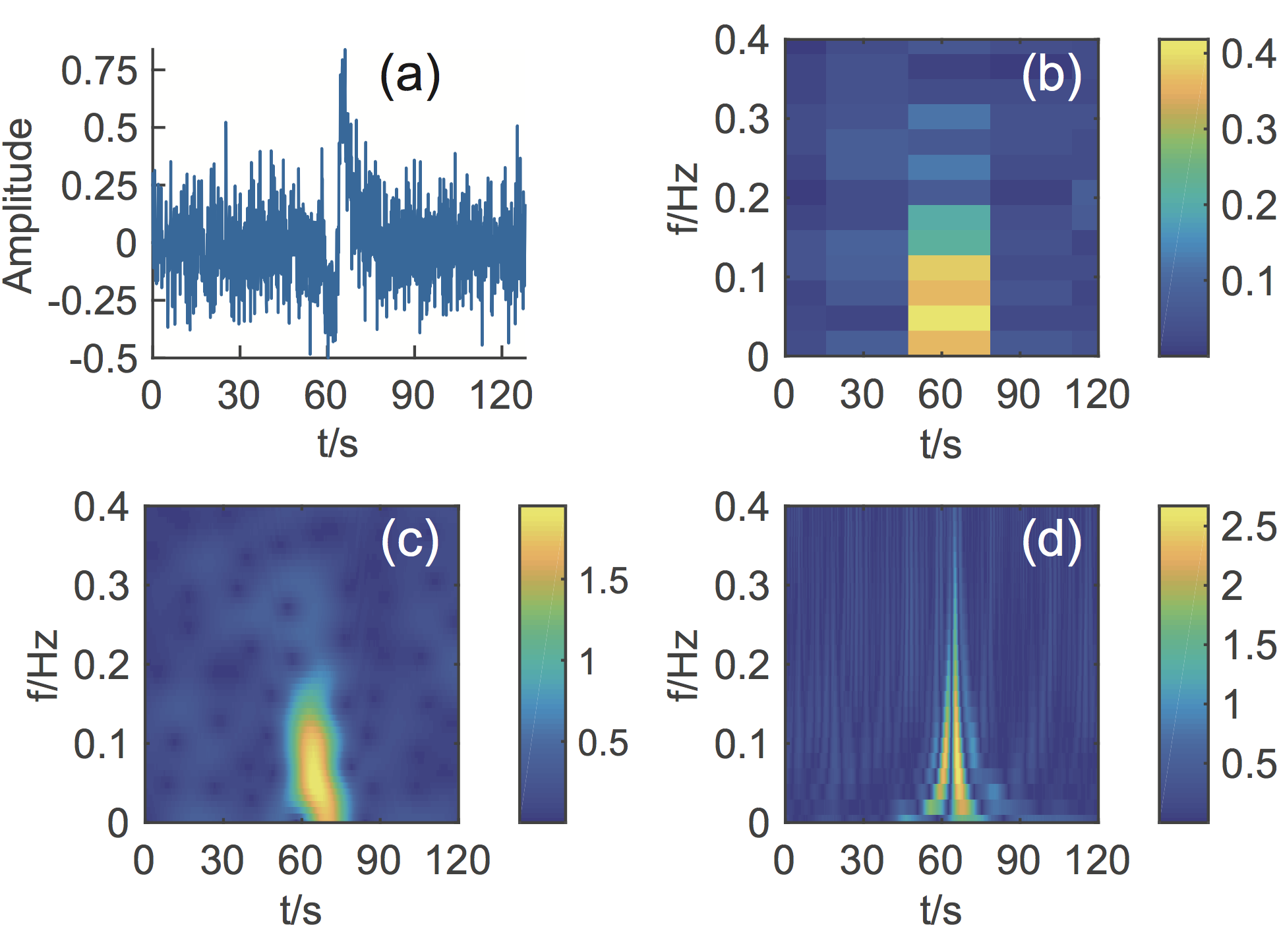} \\
\caption{ (color online) 
 (a) A signal with noise;
 (b) time-frequency analysis with our Wannier basis;
 (c) time-frequency analysis with the short-time Fourier transform;
 (d) time-frequency analysis with a wavelet method.
 }
 \label{signal}
\end{figure}

Similar to Wigner function, our method can be applied to any pair of variables which are related 
to each other by Fourier transform; 
therefore, our method has a  potential in application of time-frequency signal analysis
which is used to characterize and manipulate signals whose statistics vary in time, such as transient signals.

In the past decades, many time-frequency analysing techniques have been devised~\cite{signal}. Wavelets and short-time Fourier transform are two most prevalent methods. We compare our method with these techniques for a simulated example.  The most important goal in signal analysis is to extract the signal from the noise.  So   a testing signal is 
designed with a random noise and it is shown in Fig.\ref{signal} (a). Fig.\ref{signal} (b), (c), and (d) are results of our method, short-time Fourier transform, and wavelet, respectively.  Our method works as well as the other two  methods 
to identify the signal. However,  it is clear from Fig.\ref{signal} (b) that 
the result produced with our method is much more compact when stored on computer. 
In addition,  our method with Wannier basis has the same frequency resolution for  the whole  frequency spectrum;
in contrast, wavelet has lower frequency resolution for high frequencies. This means that our method 
should be better than wavelet when dealing with problems that require  
high frequency-resolution in regions with high frequencies.

\section{Conclusions}
We have developed a method that can map a wave function unitarily onto phase space with 
a complete set of localized Wannier functions. Our method is significantly 
improved over the von-Nuemann's method and a method in Ref.\cite{Han} with the use of 
the  L\"{o}wdin's orthogonalization. This approach is not only independent of orthogonalization order
but also more numerically efficient.  Various examples are used to  compare our method to Wigner function, 
the most popular tool used to map a wave function onto phase space. The greatest advantage of our method
over Wigner function is that our method can smooth out oscillations in wave function without losing any 
information and produce a probability distribution resembling its classical trajectory.
As a result, our method builds a better quantum-classical connection. In addition, our method 
has a great potential  in signal analysis. In the future, it will be very interesting to generalize our method
to quantum spin systems as Wigner function~\cite{TilmaPRL}. 

\section{acknowledgements}
We thank Zhigang Hu for helpful discussion. 
This work was supported by the The National Key Research and Development Program of China (Grants No. 2017YFA0303302) and the National Natural Science Foundation of China (Grants No. 11334001 and No. 11429402).

%

\end{document}